\documentclass[a4paper,11pt]{article}
\pdfoutput=1 

\usepackage{jinstpub} 
\usepackage{siunitx}

\usepackage{lineno}

\title{TCAD Simulation of Stitching for Passive CMOS Strip Detectors}

\author[a,1]{M. Baselga,\note{Corresponding author.}}
\author[b]{J-H. Arling,}
\author[b]{N. Davis,}
\author[c]{J. Dingfelder,}
\author[b,c]{I. M. Gregor,}
\author[d]{M. Hauser,}
\author[c]{F. Hügging,} 
\author[d]{K. Jakobs,}
\author[e]{M. Karagounis,} 
\author[d]{R. Koppenhöfer,}  
\author[a]{K. Kröninger,} 
\author[d]{F. Lex,}
\author[d]{U. Parzefall,} 
\author[a]{B. Sari,} 
\author[b]{S. Spannagel,} 
\author[d]{D. Sperlich,} 
\author[a]{J. Weingarten,} 
\author[d]{I. Zatocilova}

\affiliation[a]{TU Dortmund University,\\Otto-Hahn Str. 4a, Dortmund, Germany}
\affiliation[b]{Deutsches Elektronen Synchrotron (DESY),\\Notkestrasse 85, 22607 Hamburg, Germany}
\affiliation[c]{University of Bonn,\\Nussallee 12, 53115 Bonn, Germany}
\affiliation[d]{Albert-Ludwigs-Universität Freiburg,\\Hermann-Herder-Strasse 3, Freiburg, Germany}
\affiliation[d]{Fachhochschule Dortmund,\\Sonnenstraße 96, Dortmund, Germany}

\emailAdd{marta.baselga@cern.ch}

\abstract{Most of the tracking detectors for high energy particle experiments are filled with silicon detectors since they are radiation hard, they can give very small spatial resolution and they can take advantage of the silicon electronics foundries’ developments and production lines.

Strip detectors are very useful to cover large areas for tracking purposes, while consuming less power per area compared to pixel sensors. The majority of particle physics experiments use conventional silicon strip detectors fabricated in foundries that do not use stitching, relying on a very small number of foundries worldwide that can provide large amounts of strip detectors. Fabricating strip detectors in a CMOS foundry opens the possibility to use more foundries and to include active elements in the strips for future productions. For the passive CMOS strip detectors project we fabricated strip detectors in a CMOS foundry using two \SI{1}{\cm^2} reticles that are stitched together along the wafer.
The fabricated strips stitched the reticles three and five times, and it was shown that the performance of those strips is not affected by the stitching. 

This paper shows 3D TCAD simulations of the stitching area to investigate the possible effects stitching can have on the performance of the strip detectors, considering different stitching mismatches. We will show that the mismatch of stitched structures up to \SI{1}{\micro\m} does not impact the performance with TCAD simulations. 
}


\arxivnumber{2409.17749} 

\proceeding{IWORID2024\\
 30 June 2024 to 4 July 2024 \\
  Lisabon}

\begin{document}
\maketitle
\flushbottom

\section{Introduction}
\label{sec:intro}

Current tracker systems for high energy particle experiments (such as ATLAS or CMS at the LHC) are covered with semiconductor detectors. Pixelated silicon sensors give outstanding precision, but outer trackers that need to cover larger area of detectors utilise mostly strip detectors. Strip detectors give great resolution in one dimension and when stacked smartly they give precise tracking positioning. Strip silicon detectors require fewer readout channels and consume less power than the pixel layers. 

Large area strip detectors are currently being manufactured by very few foundries, limiting the availability, which can be a constraint when building future experiments. CMOS foundries have a long experience building pixelated detectors, starting with the imaging CMOS cameras and exporting the technology for high energy particle experiments. For building large area strip detectors, CMOS foundries have to stitch together several reticles, and the strip implants and metal layers have to be stitched between reticles. The impact of stitching reticles during the fabrication is unknown although the implant of the strip might be affected. Pixelated sensors avoid stitching pixel implants but strip detectors have to be stitched in the read out implant. 

One of the advantages of using a CMOS foundry for the strip production is that it has the capability to fabricate active sensors, embedding basic electronics in the strip.

LFoundry \cite{lfoundry} built passive CMOS strip detectors stitching two different \SI{1}{\cm^2} area reticles with two different strip lengths: \SI{2.1}{\cm} long and \SI{4.1}{\cm} long, having three or five reticles stitching junctions respectively. The sensors were fabricated with the \SI{150}{\nano\m} LFoundry process and used a float zone \SI{150}{\micro\m} thick, 3-\SI{5}{\kilo\ohm} resistivity wafer. They were extensively tested before and after irradiation, with laser, radioactive sources and in test beams \cite{a,b,c,d,e}. Passive CMOS strip detectors do not show any effect from stitching, giving the same results as expected from a non stitched strip detector. 

In this work we try to estimate the effect of a mismatch during stitching in TCAD (Technology computer-aided design) simulations. The simulation allows to speculate for possible mismatches between reticles although the real stitching mismatch is not fully known and it can vary from sample to sample. 

With the TCAD simulations we investigate the effect of the stitching in the strip sensors, and show that, studying possible mismatching cases up to \SI{1}{\micro\m}, stitching strips does not affect perceptively the performance of the strips. These results agree with the results shown in previous works \cite{a,b,c,d,e} where stitching shows no impact on the strip performance.

\section{Possible stitching mismatches}
\label{sec:stitching}

\begin{figure}[htbp]
\centering 
\includegraphics[width=.24\textwidth]{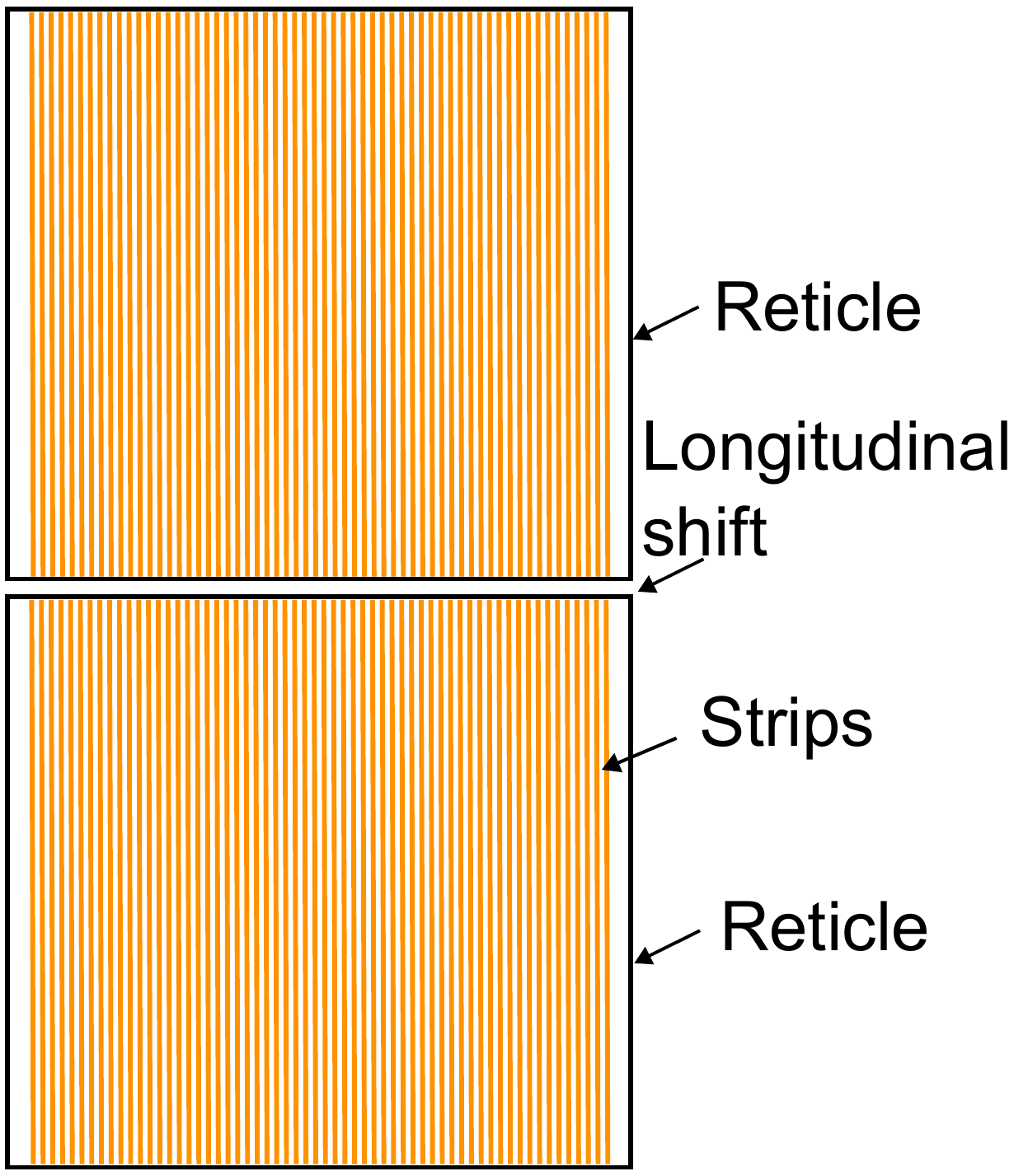}
\includegraphics[width=.24\textwidth]{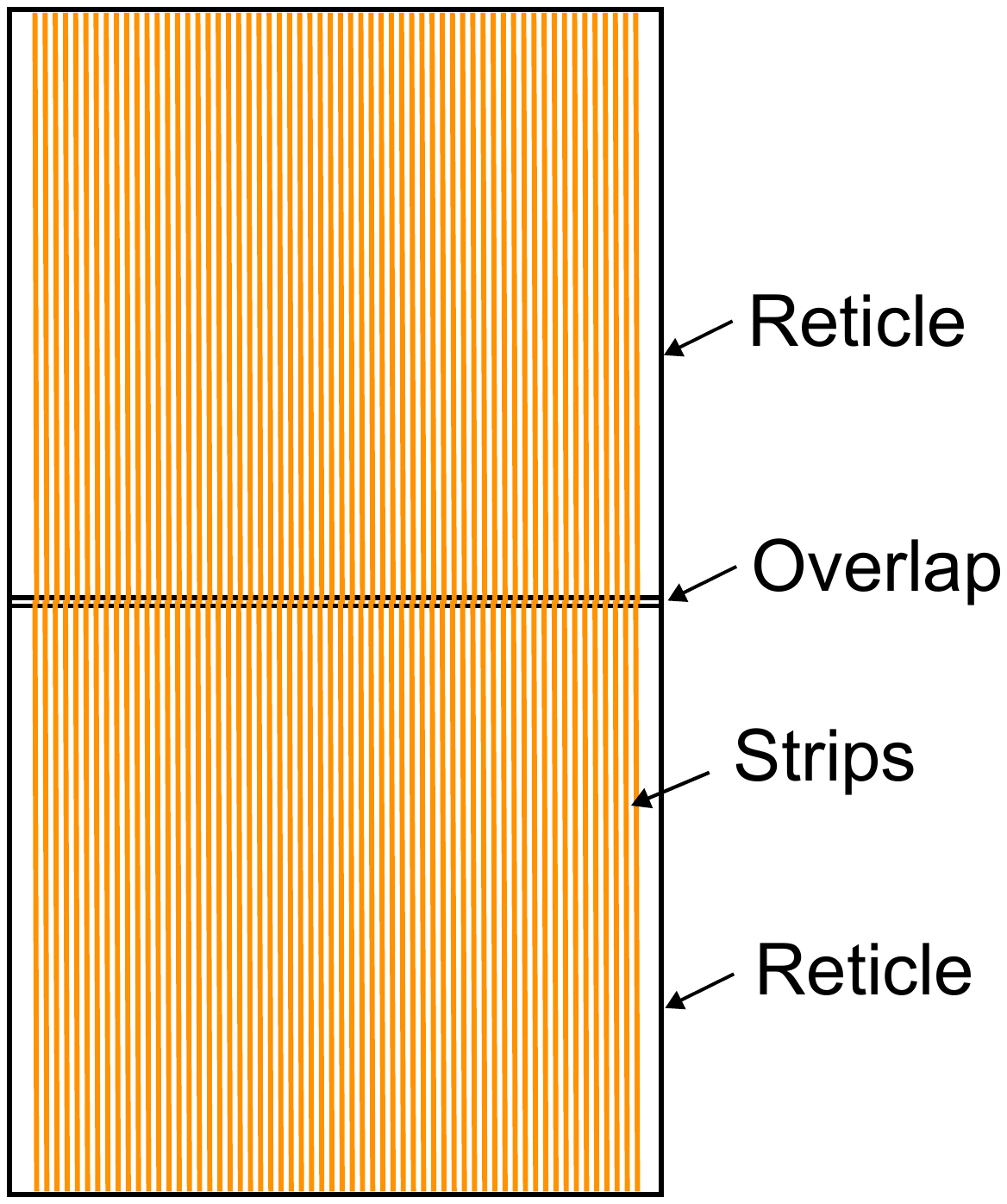}
\includegraphics[width=.24\textwidth]{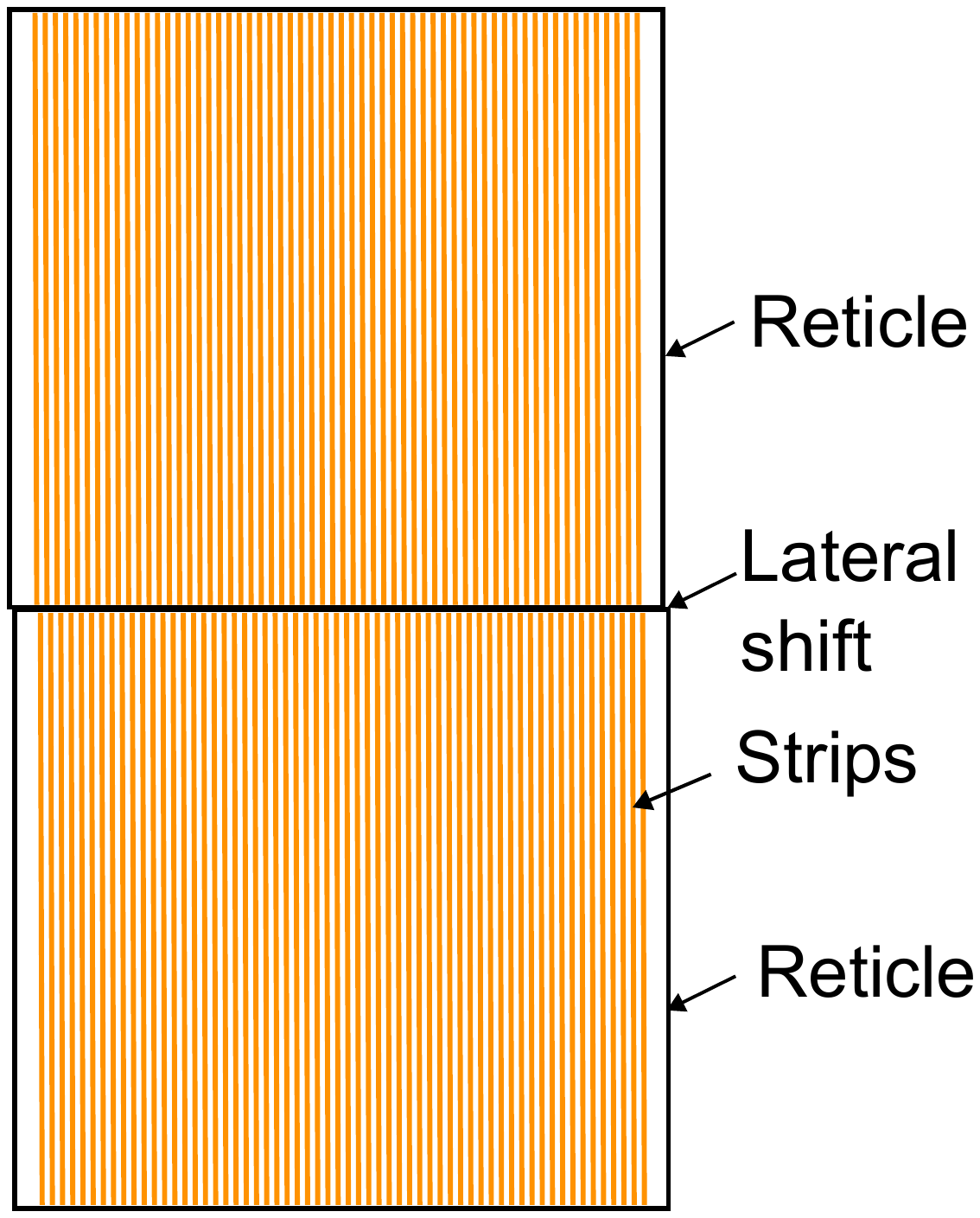}
\includegraphics[width=.24\textwidth]{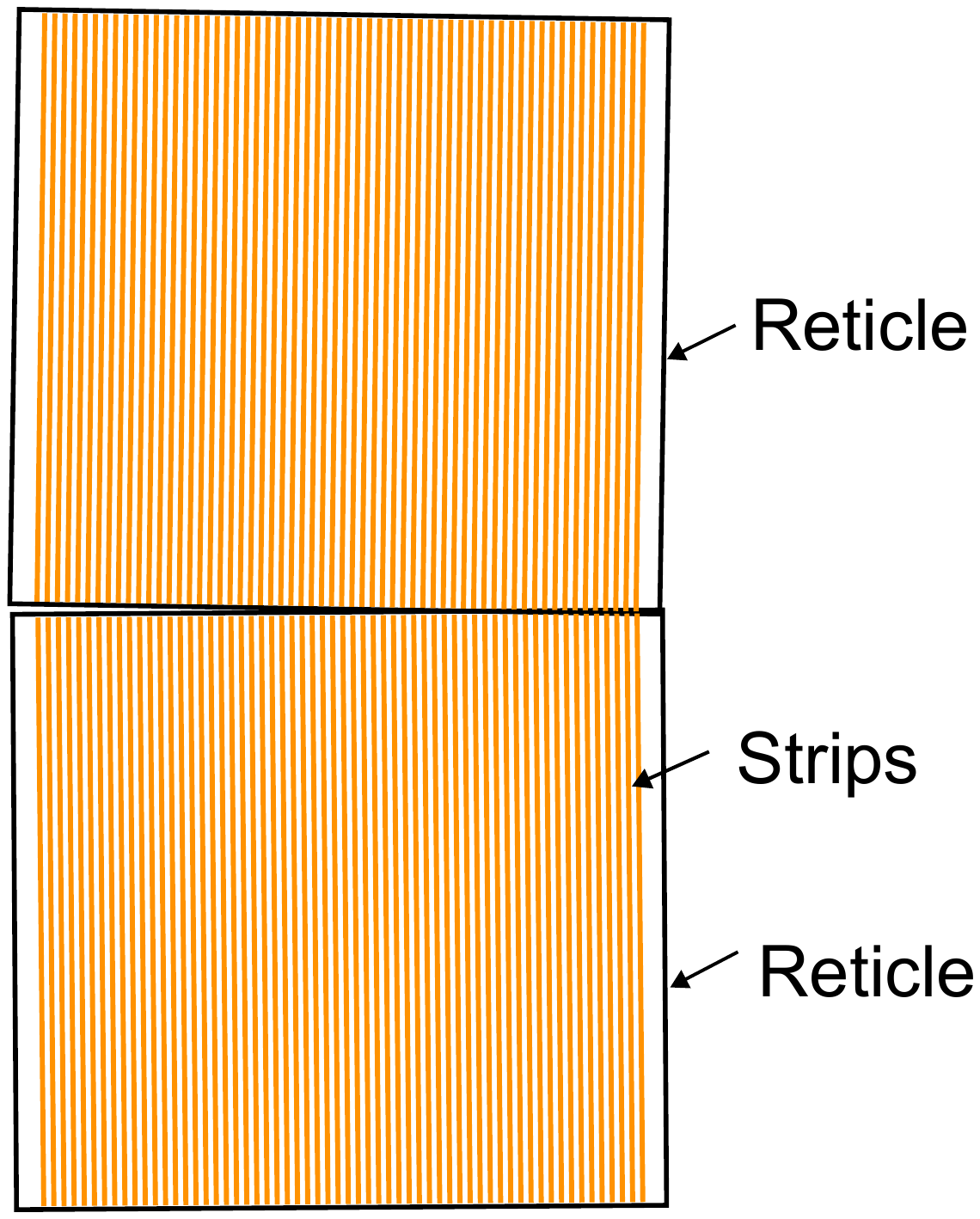}
\caption{\label{fig:stitching} Sketch of possible stitching mismatches (not to scale). From left to right: longitudinal shift, overlapped, lateral shift shift and rotation of the reticles.}
\end{figure}

When fabricating large area strip detectors in CMOS foundries, the strip implant and upper layers (metal and isolating material) will be stitched. The precision of stitching is given by the foundry process, and the alignment of the stepper motor used when doing the photolithography step for each reticle. Considering that the alignment is not perfect and some mismatch happens, the stitching can have the two implants separated with a gap (longitudinal shift), the two reticles can overlap, have a lateral shift or rotate. In reality the stitching will have a combination of all the possible mismatches, but here for simplification only the lateral and longitudinal shifted implant mismatches are considered. The overlapping mismatch will not be impacting in the photolitography, therefore it will not affect the next process steps, not being a stitching problem. The rotation mismatch will not be considered for the complexity of the possible rotation cases, although the possible result can be considered as the combination of the lateral and longitudinal shifted effect. Figure \ref{fig:stitching} shows a graphical representations of the possible stitching mismatches (not to scale). 

Here in this paper the simulation of the longitudinal and lateral shifted stitching is considered with two cases. First a  \SI{150}{\nano\m} stitching mismatch that corresponds to the LFoundry process node used for the passive CMOS strip fabrication. This stitching case gives a realistic estimate of the worst case scenario possible during production. For comparison, we simulated a \SI{1}{\micro\m} stitching, which is not realistic but tests the most extreme effects on the detector performance due to stitching misalignment, and magnifies each effect to have a better overview of the possible consequences for large stitching mismatches. For reference, a non stitched structure is also simulated.  


\section{TCAD simulations}
\label{sec:tcad}

The simulations are conducted with the Synopsys Sentaurus \cite{synopsys} TCAD software kit, which allows process and electrical simulations of the structures. The simulations of the stitching were done designing a 3D strip with a \SI{75.5}{\micro\m} pitch and \SI{150}{\micro\m} thick wafer similar to the fabricated passive CMOS strips \cite{a,c}. The simulations presented in the previous references show a good agreement with the obtained results from the fabricated sensors. The simulated strip is \SI{2}{\micro\m} long due to computational limitations, containing the stitching shift. The simulation has a strip implant of \SI{15}{\micro\m} width, the surface around the implants is filled with an STI (Shallow Trench Isolation), and the strips are isolated with p-stop structures (floating p+ implants between strips for avoiding cross talk). The backplane is also implanted with p++ impurities to propitiate a good conductivity to the metal connection.

\begin{figure}[htbp]
\centering 
\includegraphics[width=0.8\textwidth]{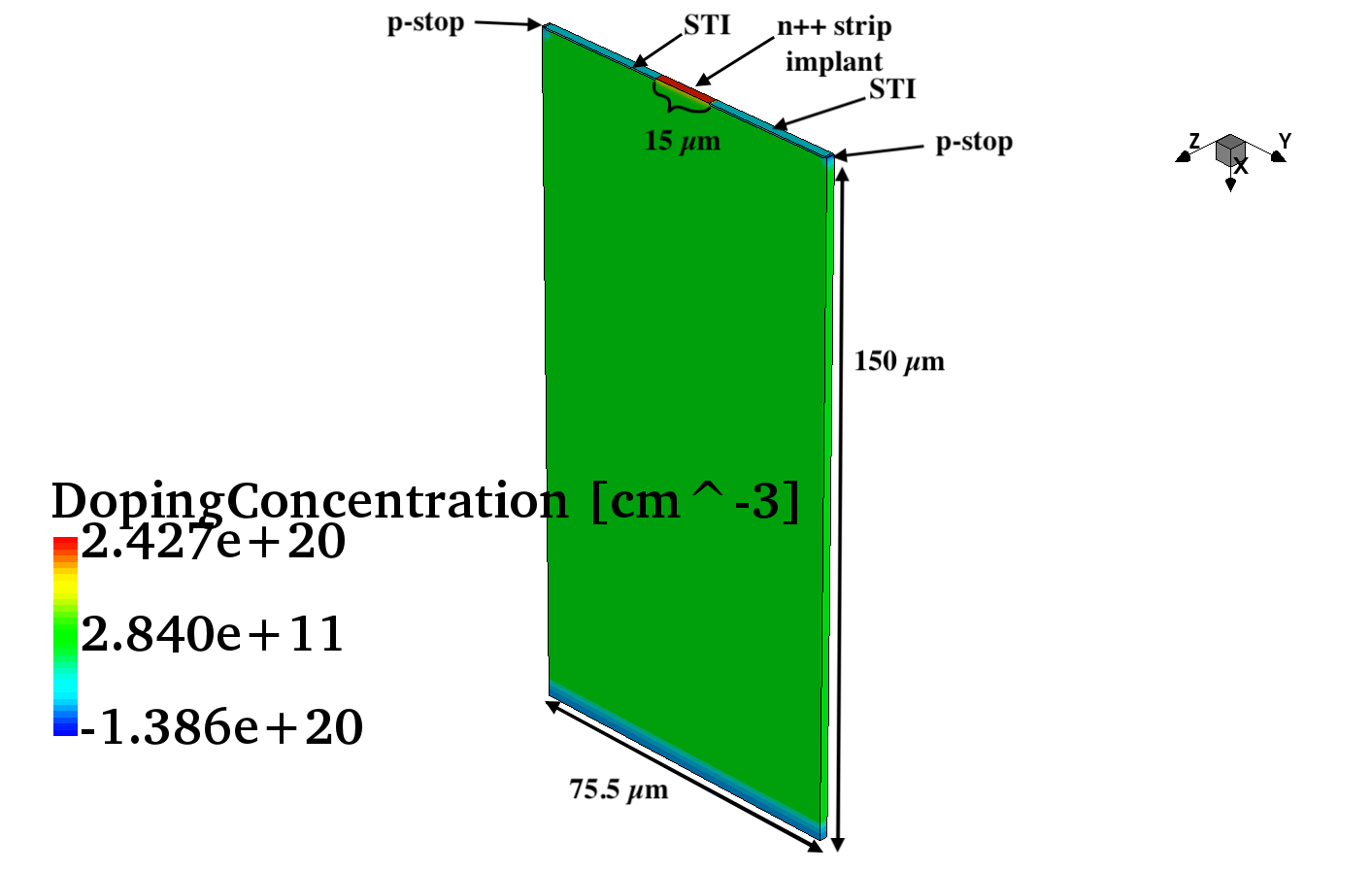}
\caption{\label{fig:doping} TCAD simulation of the doping concentration profile of the full simulated structure, with a strip length of \SI{2}{\micro\m}, pitch of \SI{75.5}{\micro\m} and \SI{150}{\micro\m} thickness.}
\end{figure}

\begin{figure}[htbp]
\centering 
\includegraphics[width=\textwidth]{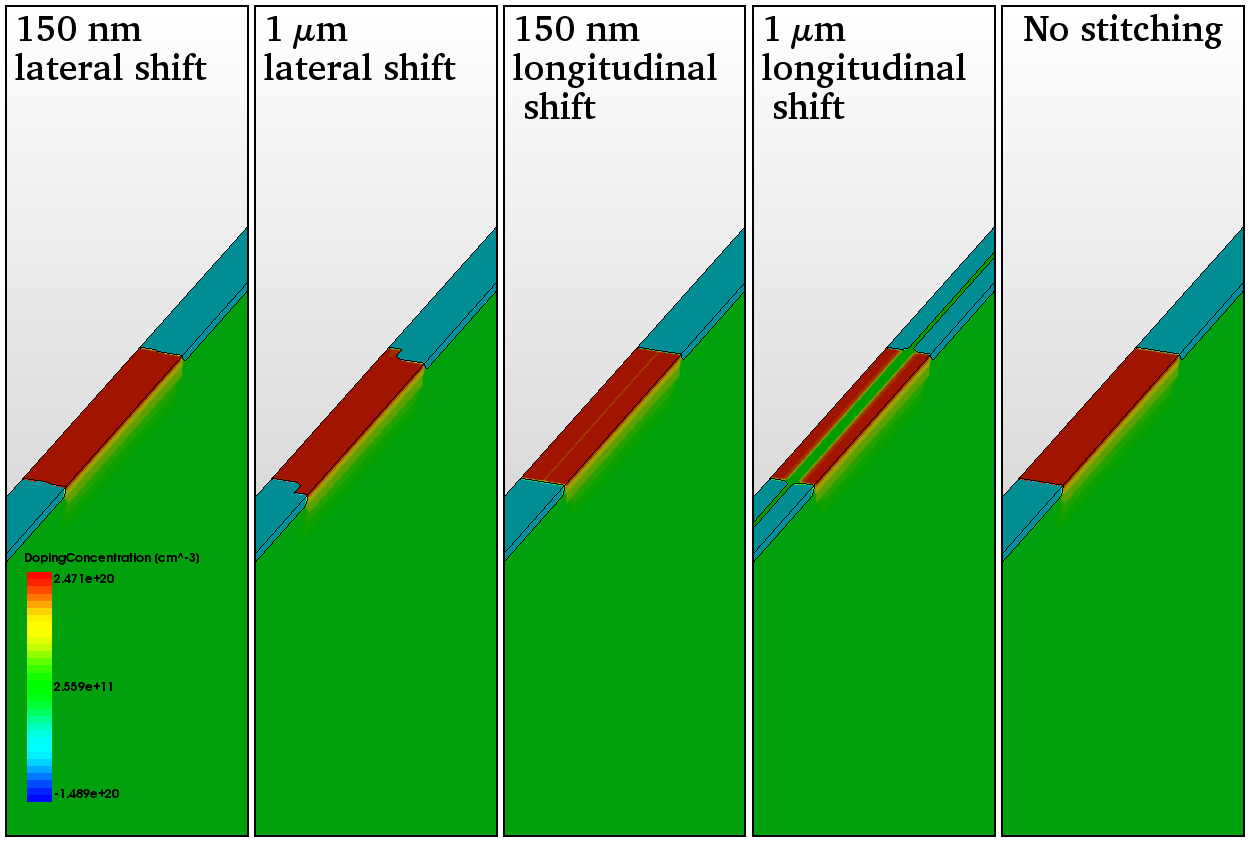}

\caption{\label{fig:doping_details} TCAD simulation of the doping concentration profile of the different studied mismatched cases, with the zoom at the strip. From left to right, lateral shift of \SI{150}{\nano\m}, lateral shift of \SI{1}{\micro\m}, longitudinal shift of \SI{150}{\nano\m}, longitudinal shift of \SI{1}{\micro\m} and with no stitching effect for comparison.}
\end{figure}

Figure \ref{fig:doping} shows the doping concentration profile of the TCAD simulations for the full structure. TCAD simulations are performed with Sentaurus Process, a software tool that can mimic the foundry fabrication steps. Figure \ref{fig:doping_details} shows the doping profile of the different studied mismatching cases in the implant strip, showing the detail of the simulated mismatches. The simulated mismatches are \SI{150}{\nano\m} and \SI{1}{\micro\m} lateral shift, \SI{150}{\nano\m} and \SI{1}{\micro\m} longitudinal shift, and the reference simulation (no stitching case).

From the simulation of the doping profile shown in Figure \ref{fig:doping_details}, the \SI{150}{\nano\m} stitching mismatch is barely visible but the \SI{1}{\micro\m} stitching mismatches are visually visible.  


\subsection{Current-Voltage Characteristics}

\begin{figure}[htbp]
\centering 
\includegraphics[width=0.47\textwidth]{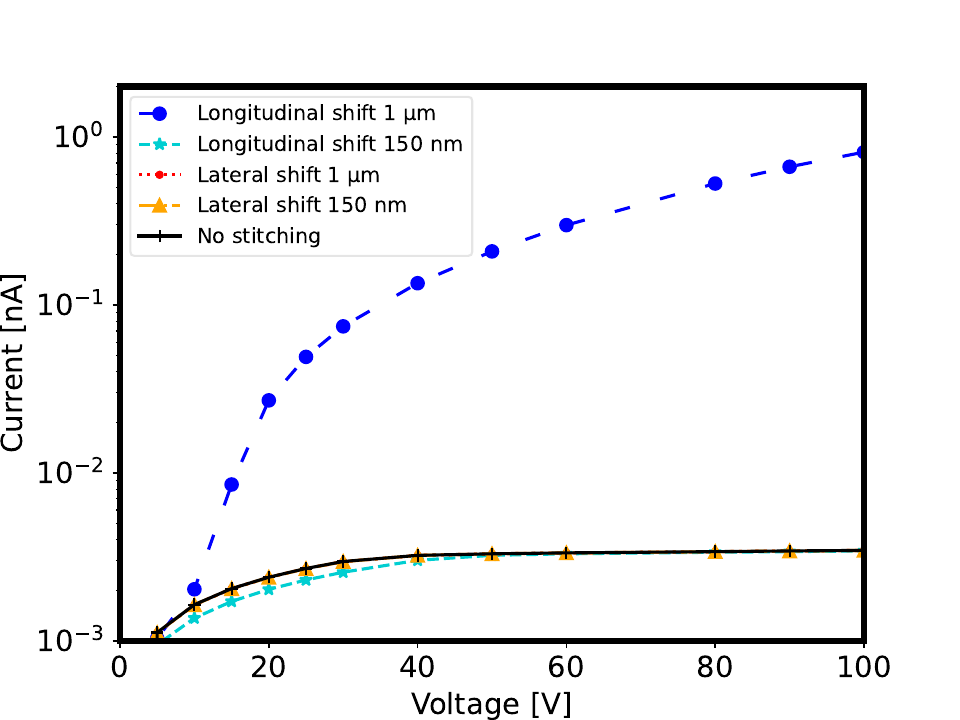}
\includegraphics[width=0.47\textwidth]{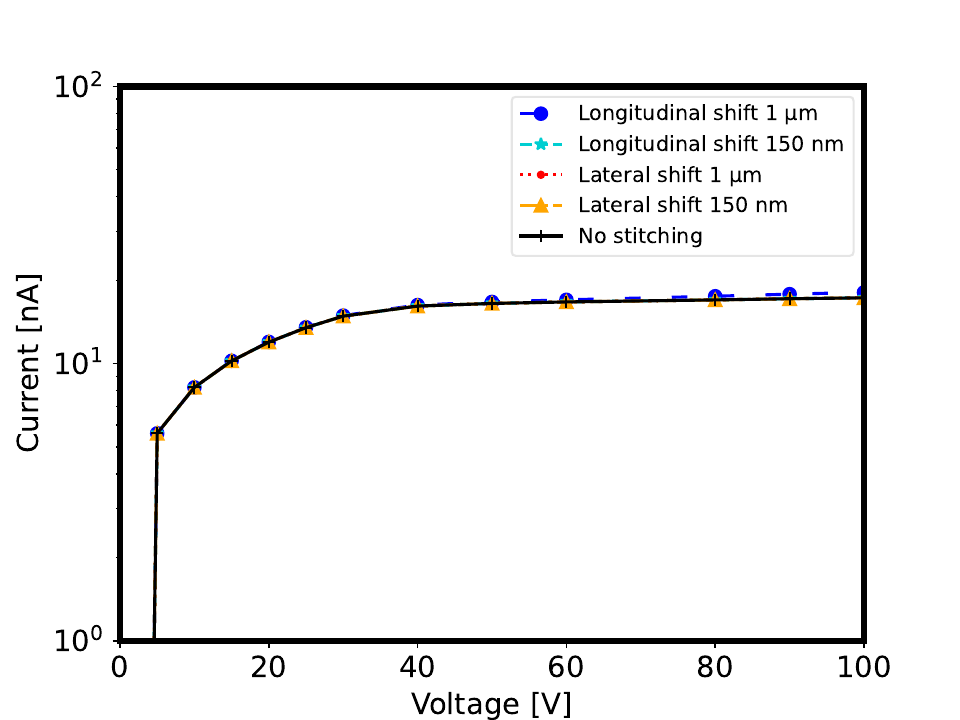}
\caption{\label{fig:iv} Current voltage curve for the studied stitching mismatches. Left plot is the IV for a \SI{2}{\micro\m} long strip simulation. Right plot is the same simulated IV mismatches as the left plot scaled for \SI{1}{\cm} long reticle.}
\end{figure}

The simulation of the current-voltage characteristic gives an indication of whether a structure leads to an earlier breakdown or to a higher leakage current, which can lead to an increase in the noise of the detector. Figure \ref{fig:iv} left plot shows the simulation of the current voltage characteristics for the studied cases up to a bias voltage of \SI{100}{\V}. Three of them show the same leakage current as the non-stitched case. while the case with the \SI{1}{\micro\m} longitudinal shift stitching shows two order of magnitude higher leakage current than the others. Nevertheless this is a \SI{2}{\micro\m} simulation, when scaled to a full reticle (for example, the \SI{1}{\cm} reticle used for the built LFoundry passive strip detectors) the increase of the leakage current is rather low, as can be appreciated in Figure \ref{fig:iv} right plot. Figure \ref{fig:iv} right plot shows the current voltage characteristics simulation for the different \SI{2}{\micro\m} mismatches scaled for \SI{1}{\cm} long reticle.

\subsection{Capacitance-Voltage Characteristics}

\begin{figure}[htbp]
\centering 
\includegraphics[width=0.47\textwidth]{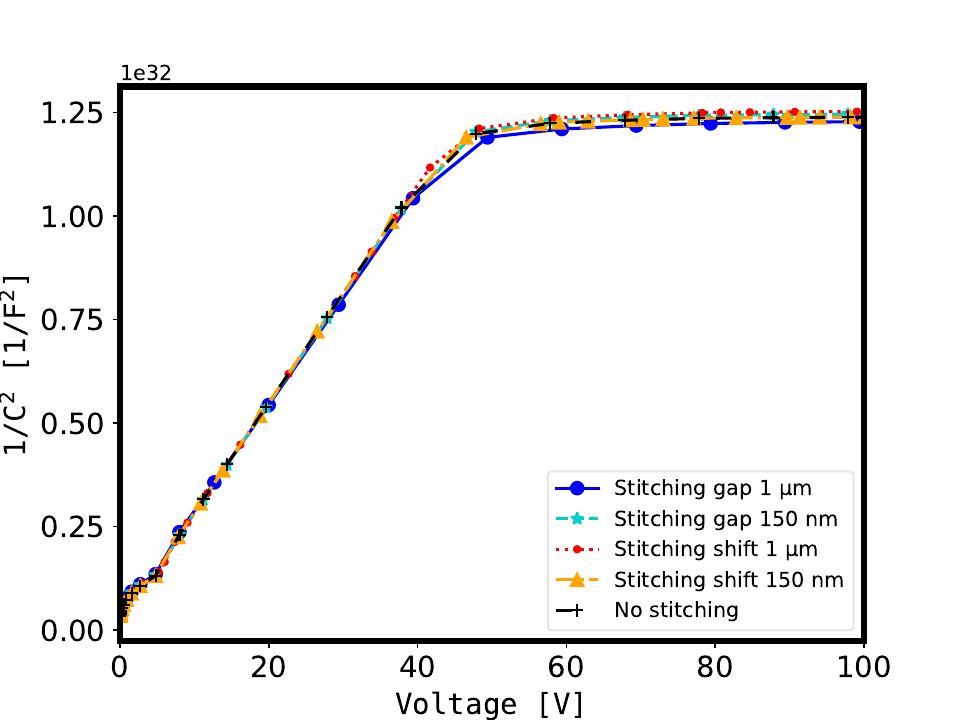}
\caption{\label{fig:cv} Capacitance voltage curve for the studied mismatches.}
\end{figure}

TCAD allows also to simulate the capacitance voltage characteristics of the devices. In this case, a test pulse with frequency of \SI{1}{\kilo\Hz} is applied to the simulation. Figure \ref{fig:cv} shows the capacitance voltage characteristics for the five different studied cases.

The simulation does not reveal any difference between the mismatches, giving a good agreement with the results shown for the built devices. Therefore, the simulations indicates that the stitching mismatch does not affect the depletion of the sensor.

\subsection{Electric field analysis}

\begin{figure}[htbp]
\centering 
\includegraphics[width=\textwidth]{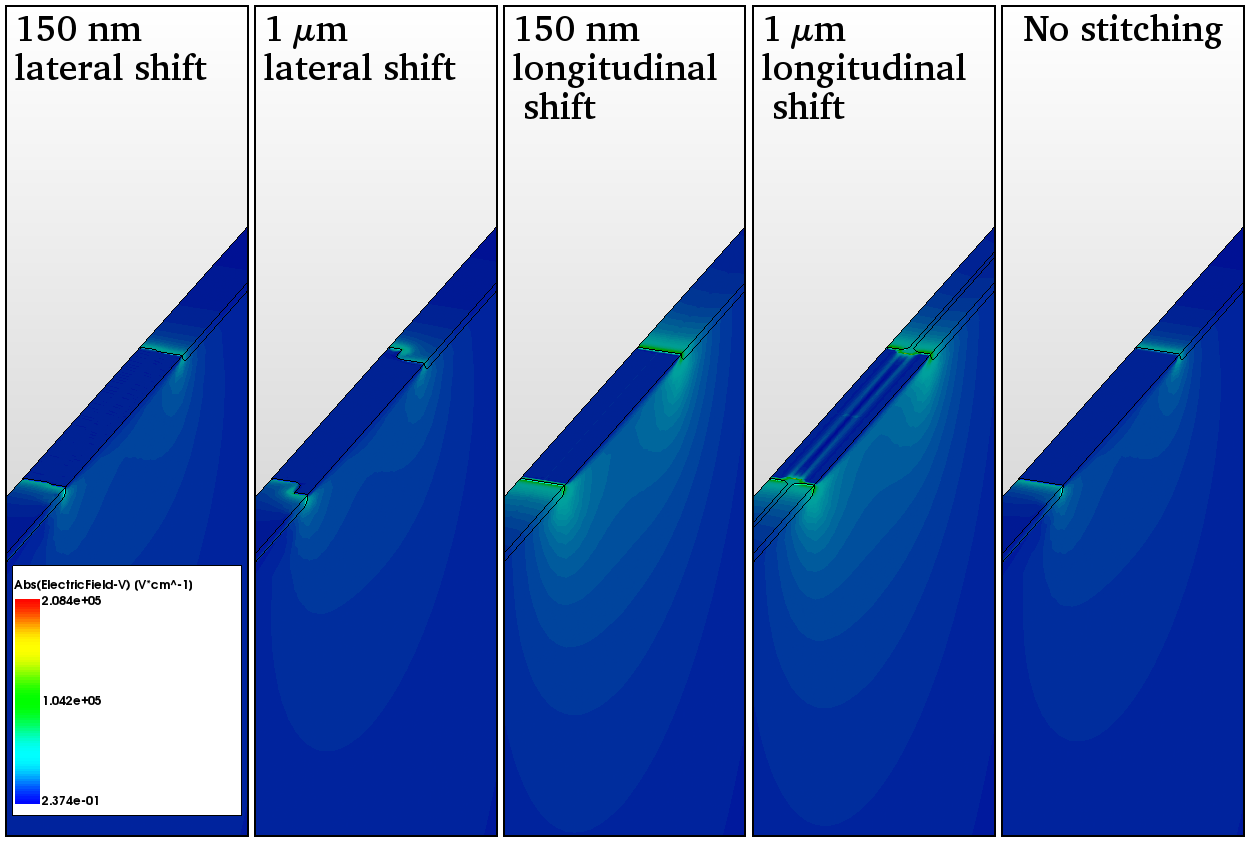}

\caption{\label{fig:E_all} Detail of the electric field of the different mismatches in the strip implant region. From left to right, lateral shift of \SI{150}{\nano\m}, lateral shift of \SI{1}{\micro\m}, longitudinal shift of \SI{150}{\nano\m}, longitudinal shift of \SI{1}{\micro\m} and with no stitching for comparison.}
\end{figure}

Synopsys Sentaurus can simulate the electric field of the device. Figure \ref{fig:E_all} shows the electric field at \SI{100}{\V} for all the studied mismatching cases. 

The electric fields for all the studied cases are almost the same, except the longitudinal shift stitching of \SI{1}{\micro\m} that has a higher electric field in the stitching gap, although not as high as the electric field of the junction, showing that the electric field is not highly affected from the stitching mismatches. 

\subsection{Charge collection comparison}

\begin{figure}[htbp]
\centering 
\includegraphics[width=0.47\textwidth]{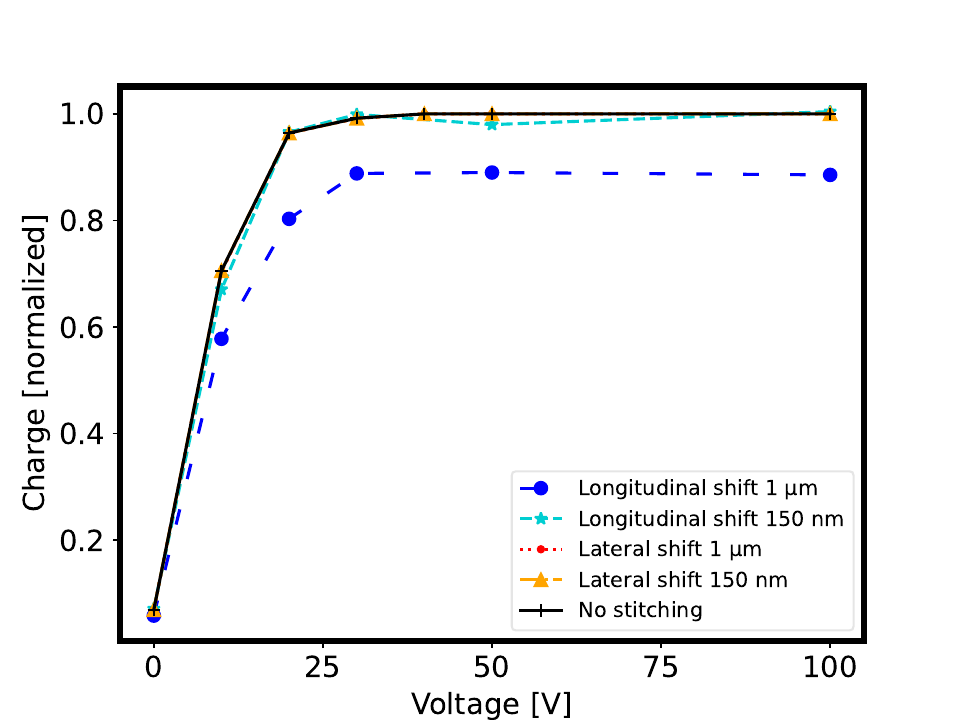}
\includegraphics[width=0.47\textwidth]{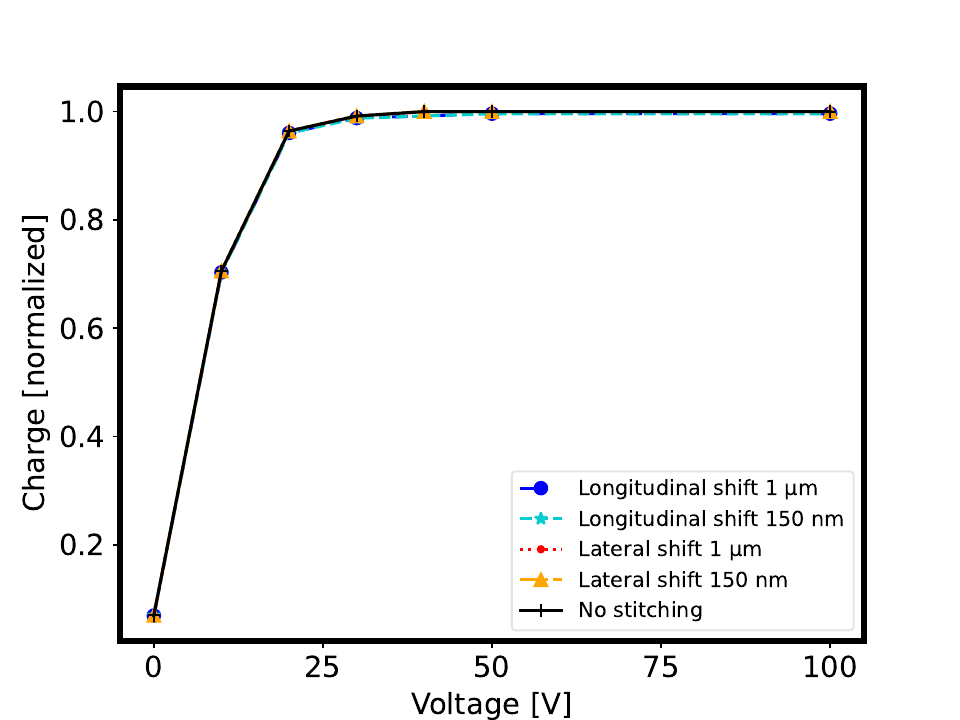}
\caption{\label{fig:charge} Charge collection efficiency as a function of the bias voltage curve for the simulated structures. Left plot is the simulation for a \SI{2}{\micro\m} long strip and the right plot is scaled to \SI{1}{\cm} long reticle strip. }
\end{figure}

For simulating the charge collection, a charge mimicking the deposition of a MIP (Minimum Ionizing Particle) is created in the central position, in the middle of the strip and goes through the full bulk thickness (\SI{150}{\micro\m}), and then integrating the charge over \SI{25}{\nano\s}. Figure \ref{fig:charge} left plot shows the collected charge for different voltages for the \SI{2}{\micro\m} long strip from the simulation. Only the collected charge for the \SI{1}{\micro\m} longitudinal stitching shift shows a significant reduction with respect to the non-stitched case. 
Nevertheless, when scaling the \SI{2}{\micro\m} simulation to the \SI{1}{\cm} reticle length the decrease of the charge for the wider stitching gap is negligible (shown in Figure \ref{fig:charge} right plot).

\section{Conclusions}
\label{sec:conclusions}

In this paper we studied the possible effect of lateral and longitudinal shifts in the stitching of strip sensors with TCAD simulations. 
TCAD simulations of the \SI{150}{\nano\m} stitching mismatches do not show any impact to the performance of the strip detector. The simulations shown in this work are only for \SI{2}{\micro\m} long strip, therefore when scaling with a full length strip the effects shown are negligible. These results concur with the previous measurements conducted in the lab with radioactive source, transient current techniques and in test beams with electrons, that the fabricated detectors do not show any effect of the stitching. 


\acknowledgments

This work has been partially funded by the BMBF
grant Verbundproject 05H2021 - R\&D
DETEKTOREN (Neue Trackingtechnologien): Entwicklung von
aktiven und passiven mikrostrukturierten CMOS-Sensoren.

\end{document}